\documentclass[conference]{IEEEtran}
\usepackage{ltexpprt}
\usepackage{amssymb}
\usepackage{epsfig} 
\usepackage{subfigure}
\usepackage{xspace}

\usepackage{algorithm}
\usepackage{algorithmic}

\newcommand{\sna}{Social Network Analysis\xspace}

\newcommand{\pr}{PageRank\xspace}
\newcommand{\lr}{logistic regression\xspace}

\newcommand{\skr}{Supervised Kemeny Ranking\xspace}

\newcommand{\lk}{Local Kemenization\xspace}

\newcommand{\pk}{Approximate Kemeny\xspace}
\newcommand{\pktk}{AK TopK\xspace}
\newcommand{\suppktk}{Supervised AK TopK\xspace}

\newcommand{\ak}{Approximate Kemeny\xspace}

\newcommand{\drop}[1]{}

\newcommand{\prem}[1]{{\bf [Prem:{\em {#1}}]}}

\newcommand{\ap}{Average Precision\xspace}
\newcommand{\bordaunsup}{Borda\xspace}
\newcommand{\bordasup}{Supervised Borda\xspace}

\newcommand{\lkunsupborda}{Unsupervised, Total Orderings\xspace}
\newcommand{\lksupborda}{Supervised, Total Orderings\xspace}
\newcommand{\lktkunsupbordasup}{Unsupervised, Partial Orderings\xspace}

\newcommand{\borda}{Borda\xspace}
\newcommand{\supborda}{Supervised Borda\xspace}

\newcommand{\ecc}{Extended Condorcet Criterion\xspace}

\newcommand{\rtgindegree}{Distinct Past Retweets\xspace}
\newcommand{\rtgwtdindegree}{Past Retweets\xspace}
\newcommand{\fgindegree}{Followers\xspace}
\newcommand{\fgoutdegree}{Friends\xspace}
\newcommand{\rtgoutdegree}{People Retweeted\xspace}
\newcommand{\rtgwtdoutdegree}{Retweets Made\xspace}
\newcommand{\cgoutdegree}{People Mentioned\xspace}
\newcommand{\cgwtdoutdegree}{Mentions Made\xspace}
\newcommand{\cgindegree}{Distinct Mentions Received\xspace}
\newcommand{\cgwtdindegree}{Mentions Received\xspace}

\newcommand{\rtgpagerank}{Retweet Pagerank\xspace}
\newcommand{\fgpagerank}{Follower Pagerank\xspace}
\newcommand{\cgpagerank}{Mention Pagerank\xspace}

\newcommand{\mg}{Mention Graph\xspace}

\ifCLASSINFOpdf
\else
\fi
\begin{document}
%
\title{Supervised Rank Aggregation for Predicting Influence in Networks}

\author{Karthik~Subbian and Prem Melville\\
IBM T.J. Watson Research Center, Yorktown~Heights,~NY~10598\\
Email: \{ksubbian,pmelvil\}@us.ibm.com}

%


\maketitle

\begin{abstract} 
Much work in Social Network Analysis has focused
on the identification of the most important actors in
a social network. This has resulted in several measures of
influence and authority\drop{, centrality or prestige}. 
While most of such sociometrics (e.g., PageRank) are driven by intuitions based on an
actor's location in a network, asking for the ``most influential'' actors in itself is an ill-posed question, unless it is put in context with a specific measurable task. 
Constructing a predictive task of interest in a given domain provides a
mechanism to quantitatively compare different measures of
influence. Furthermore, when we know what type of
actionable insight to gather, we need not rely on a single
network centrality measure. A combination of measures is more
likely to capture various aspects of the social network that
are predictive and beneficial for the task. Towards this end, we propose an approach to supervised rank aggregation, driven by techniques from Social Choice Theory. We illustrate the effectiveness of this method through experiments on Twitter and citation networks.
\end{abstract}

\drop{
we exploit rank aggregation techniques from Social Choice Theory, and extend them to supervised settings. We illustrate the effectiveness of these methods through a case study on a data set of 40 million Twitter users, where we study influence in the context of predicting when users will be rebroadcast. a case study on a data set of 40 million Twitter users, where we study influence in the context of predicting when users will be rebroadcast.
}


%
\IEEEpeerreviewmaketitle


%

\section{Introduction}
The rise of Social Media, with its focus on user-generated content and social networks, 
has brought the study of authority and influence in networks to the forefront. For companies and other public entities, identifying and engaging with influential authors in social media is critical, since any opinions they express could rapidly spread far and wide. 
\drop{Even for consumers of information, identifying influential sources is becoming increasing crucial.} For users, when presented with a vast amount of content relevant to a topic of interest, ordering content by the source's authority or influence assists in information triage, thus overcoming the ever-increasing information overload.

Following this need, there has been a spate of recent work studying influence and the diffusion of information in social networks~\cite{bakshy:ec09,goetz:icwsm09,kossinets:kdd08}. While these works are important in furthering our understanding of the dynamics of communication in networks, they do not directly give us measures of influence and authority in social media. On the other hand, there has been much work in the field of Social Network Analysis, from the 1930's~\cite{moreno:book34} onwards, that has focused explicitly on sociometry, including quantitative  measures of influence, authority, centrality or prestige. These measures are heuristics usually  based on intuitive notions such as access and control over resources, or brokerage of information~\cite{knoke:ana83}; and has yielded measures such as Degree
Centrality, Eigenvector Centrality and Betweeness Centrality~\cite{wasserman:book94}. 

\drop{To our knowledge, the task of identifying influencers in a predictive framework using various socio-metric measures is not studied so far.} In this paper, we address the problem of identifying influence by posing it as a predictive task. In particular, we compare different measures of influence on their ability to accurately predict which users in Twitter will be virally rebroadcast (\emph{retweeted}) in the near future. Formulating a concrete predictive task, such as this, allows us to quantitatively compare the efficacy of different measures of influence. 

In addition to evaluating individual measures of influence, such as Degree Centrality and PageRank, we propose combining them to produce a more accurate measure of influence. 
Given that each measure produces an ordering of elements, we can leverage rank aggregation techniques from Social Choice Theory, such as Borda~\cite{borda1781} and Kemeny optimal rank aggregation~\cite{kemeny59}. These classical techniques were designed to combine rankings to ensure \emph{fairness} amongst voters and not to maximize performance on a predictive task; and as such are \emph{unsupervised}. In this paper, we introduce \skr in order to aggregate individual rankings for the task of predicting influence in networks. We demonstrate the effectiveness of our approach in a case study of 40 million Twitter accounts; and we further corroborate these results in a study of publication citation networks. 

\drop{
In this paper, we propose a novel approach, \skr , and demonstrate its effectiveness on the two case studies. Firtly, we demonstrate the effectiveness of our \skr method in a case study of 40 million Twitter accounts. Further, we substantiate the efficacy of our method using a second case study, citation network data from KDD Cup 2003. The predictive task in the second case study is to predict the influential paper, where the ground truth in this case is the number of actual downloads occured for the paper during the test time period. In addition, we compare our method with prior art like Local Keminization \cite{dwork:www01} and SVMRank \cite{joachims:06} and show its effectiveness in terms of Area Under the Curve (AUC) and Mean Average Precision (MAP) measures.
}

\drop{\prem{Mention citation networks case study.} In particular, we show the merits of supervised locally-optimal order-based rank aggregation. 
\prem{Maybe we should pitch this stronger and more focused, such as ``we propose a novel approach, Supervised Kemeny Aggregation, and demonstrate its effectiveness...''}}

In this paper, we make the following key contributions: (1) We propose a predictive, rather than a heuristic, perspective of influence, by formulating measurable predictive tasks.
(2) We combine ideas from Sociometry and Social Choice Theory in novel ways.   
(3) We present a new approach to supervised rank aggregation. 
(4) We show the effectiveness of our approach on real-world network data.
(5) We demonstrate that our approach is significantly better than current practice and other \drop{more sophisticated} baselines that we devised.

\section{Data Set and Task Definition}
\label{sec:data}
Our primary study was based on the Twitter discussion around Pepsi. What piqued our interest in
Twitter and the role of influencers was the infamous sexist iPhone app called ``AMP UP B4 U SCORE''. An avalanche of Twitter users slammed the app ultimately leading to an apology from Pepsi. In this study, we found that the influence of twitter users heavily depends upon the number of rebroadcasts of his/her messages to millions of other users. In the context of Twitter, this suggests that a useful task would be to predict which twitterers will be significantly rebroadcast via retweets.

One obvious indicator of influence could be the number of followers a user has (in-degree of the Follower Graph). However, many users follow 100K or more users and therefore this may not be sufficient indication of influence. For this reason, we consider two alternatives, the Retweet Graph and the Mention Graph, where edges correspond to retweets and mentions of users in the past. We generate two versions of both the Retweet  and the Mention Graph, one collapsing all repeat connections from the same user \textit{i} to the user \textit{k} into just one edge. The second version uses the number of retweets/mentions as edge weights. For our influence measures (rankings) we use in-degree, out-degree and PageRanks\drop{~\cite{brin:cn98}} (with a damping factor of 0.85).
In addition to degree and eigenvector centralities, there are other important socio-metrics
based on the paths between vertices like, Closeness and Betweeness Centrality. 
We exclude them, as they come at the prohibitive computational cost of calculating all-pairs shortest paths in a graph $(O(V^3)).$\footnote{In related work, we have been working on a scalable algorithm for computing Betweeness Centrality, exploiting hierarchical parallelization.}



We extracted the data\footnote{We will make all our data publicly available.} to generate these graphs over a two week period from 11/11/09 to 11/26/09. This gives a Follower Graph with 40 million nodes (users) and 1.1 billion edges. We used the socio-metrics computed from these graphs to predict which users will have viral outbursts of retweets in the following week. We compare these predictions with the actual amount of retweets in the following week. For the purposes of testing, we monitored all retweets of a set of 9,625 users. This is the set we use for the train-test splits in our experiments.

\drop{We construct our prediction task from our data by dividing users in our test period into two classes –-- people who have been retweeted more than 100 times (approximately 10\% of the maximum number of retweets within a week) and those who have not.} We construct our prediction task from our data by dividing users in our test period into two classes -- people who have been retweeted more than a threshold and below. In our data set, we selected ~10\% of the maximum number of retweets within a week as the threshold (100 retweets). We treat this as a binary classification problem, where the ranking produced by each measure is used to predict the potential for viral retweeting in the test time period. Since we are primarily concerned with how well these measures perform at ranking users, we compare the area under the ROC curve (AUC) based on using each measure \cite{faw:prl06}. For some applications it is more important to correctly rank relevant elements at the top of list, which we also measure by Average Precision (AP) for the top $k$ users~\cite{yates:99}.

\drop{
In addition, we also measure the effectiveness of these measures in terms of their positive predictive value or Mean Average Precision (MAP) \cite{yates:99} for the top K users. 
}

We compared all measures of influence averaged over 20 trials of random stratified samples of 80\% of the users (see Table \ref{table:measures}). We find that 9 of the 13 individual measures by themselves are quite effective at ranking the top potentially viral twitterers with an AUC \(> 80\%\). Not surprisingly, the number of times that someone has been retweeted in the recent past produces very good rankings -- based on AUC and Average Precision. The number of followers and the number of people mentioned also produce reasonably good rankings in terms of AUC and Average Precision respectively.\footnote{Despite its popularity, PageRank does not perform as well as other measures.} However the Spearman rank correlation between recent past retweets and followers is not very high (0.43), suggesting that there are multiple forces at work here. \drop{ This fact can be clearly seen in Fig. \ref{fig:viral-roc}, where we compare the composite ranking to the best individual measure from each graph. As you can see, the best measures on each of the individuals graphs perform well in different operating regions. However, our proposed method, \skr combines these individual rankings to produce a overall better composite influence measure, that performs well across all operating regions. In addition, the ranking results presented in Table \ref{table:ablation} show that removing measures from any of the three graphs diminishes
our ability to identify viral potential.} This underscores the fact that each aspect (network of followers, diffusion of
past retweets, and interactions through replies and mentions) contributes to one's potential to reach a large audience.
By focusing on selecting a single centrality measure to capture influence we would miss out on the opportunity to more 
precisely detect potential viral users.

\drop{
the total number of times that someone has been retweeted (Weighted Retweet Graph Indegree) and the number of persons retweeted in the recent past (Retweet Graph Indegree) produces very good ranking--based on AUC and MAP metrics. In the follower graph, the number of followers (Follower Graph Indegree) also produces a reasonably good ranking. However the Spearman rank correlation between recent past retweets and followers is not very high (0.43), suggesting that there are multiple forces at work here. This leads us to consider different methods of combining these individual rankings to produce a better composite influence measure.
}

\begin{table}
\scriptsize
\begin{center}
\begin{tabular}{|p{2.8cm}|p{2.7cm}|p{0.7cm}|p{0.7cm}|}
\hline
\textbf{Measure}&\textbf{Definition}&\textbf{AUC}&\textbf{AP}\\
\hline
\hline
\fgindegree     &Follower Graph Indegree&88.18&0.4366\\
\fgoutdegree    &Follower Graph Outdegree&76.03&0.2821\\
\fgpagerank     &Follower Graph Pagerank&85.77&0.4397\\
\hline
\rtgindegree    &Retweet Graph Indegree&90.17&0.7246\\
\rtgoutdegree   &Retweet Graph Outdegree&87.04&0.3976\\
\rtgpagerank &Retweet Graph Pagerank&88.38&0.5135\\
\rtgwtdindegree &Wtd. Retweet Indegree&90.18&0.7406\\
\rtgwtdoutdegree &Wtd. Retweet Outdegree&86.80&0.4707\\
\hline
\cgindegree     & \mg Indegree&60.71&0.5690\\
\cgoutdegree    & \mg Outdegree&86.11&0.5923\\
\cgpagerank     & \mg Pagerank&70.43&0.3631\\
\cgwtdindegree  & Wtd. Mention Indegree&60.53&0.2737\\ 
\cgwtdoutdegree & Wtd. Mention Outdegree&84.69&0.2895\\
\hline
\end{tabular}
\end{center}
\caption{\label{table:measures} Comparing ranking measures for identifying viral potential, in terms of AUC(\%) and Average Precision@100.}
\end{table}

\section{Rank Aggregation}
\label{sec:rank}
\drop{It is evident from Fig. \ref{fig:viral-roc} and Table. \ref{table:ablation} that the individual socio-metrics often fail to capture all the critical factors that are relevant for  predicting influence in networks. }As each socio-metric captures only some aspect of the user's influence in the network, it is beneficial to combine them in order to more accurately identify influencers. One straightforward approach to combining individual measures is to use them as inputs to a classifier, such as logistic regression, which can be trained to predict the target variable (e.g., future retweets) on historical or held-out data. However, given that the individual influence measures produce an ordering of elements and not just a point-wise score, we can, instead leverage approaches to aggregating rankings for better results. The problem of rank aggregation or preference aggregation has been extensively studied in Social Choice Theory, where there is no \emph{ground truth} ranking, and as such are unsupervised. In this section, we explain the necessary background for appreciating our proposed method \skr, which is a supervised order-based aggregation technique, that can be trained based on the ground-truth ordering of a subset of elements.

\drop{In this paper, we introduce a supervised order-based aggregation technique, that can be trained based on the ground-truth ordering of a subset of elements.}

{\bf The Rank Aggregation Task:}
Let us begin by formally defining the task of rank aggregation. Given a set of entities $S$, let $V$ be a subset of $S$; and assume that there is a total ordering among entities in $V$. We are given $r$ individual rankers ${\tau}_1,...,{\tau}_r$ who specify their order preferences of the $m$ candidates, where $m$ is size of $V$, i.e., $\tau _i  = [d_1 ,...,d_m ],i = 1,...,r,{\rm if }d_1  > ... > d_m,d_j  \in V,j = 1,...,m$. If $d_i$ is preferred over $d_j$ we denote that by $d_i > d_j$. Rank aggregation function $\psi$ takes input orderings from $r$ rankers and gives $\tau$, which is an aggregated ranking order. If $V$ equals $S$, then $\tau$ is called a {\em full list} (total ordering), otherwise it is called a {\em partial list} (partial ordering). 

All commonly-used rank aggregation methods, satisfy one or more of the following desirable $goodness$ properties: Unanimity, Non-dictatorial Criterion, Neutrality, Consistency, Condorcet Criterion and Extended Condorcet Criterion (ECC)~\cite{arrow63}. We will primarily focus on ECC, defined below: 

\begin{Definition}
\label{def:ecc}
The Extended Condorcet Criterion~\cite{truchon98} requires that if there is any partition $\{C,R\}$ of $S$, such that for any $d_i \in C$ and $d_j \in R$ a majority of rankers prefer $d_i$ to $d_j$, then the aggregate ranking $\tau$ should prefer $d_i$ to $d_j$.
\end{Definition}

The ECC property is highly preferred in our domains, as it eliminates the possibility of inferior candidates being introduced strategically in order to manipulate the choice between superior candidates. In other words, it offers the property of Independence of Irrelevant Alternatives. Additionally, ECC is a relaxed form of Kemeny optimal aggregation (defined below), where the partition $C$ and $R$ are arranged in the ``true'' order, but not necessarily the elements within partitions $C$ and $R$. In addition to the desirable theoretical properties, ECC proves to be very valuable in ranking in practice, as we will demonstrate in our experiments.

\drop{For our discussion, we will focus on ECC property, since it ensures that the majority winners are always on the top of the aggregated list. ECC is defined as	following. Split the entities into two partitions, $Q$ and $R$. If for all $d_i \in Q$ and $d_j \in R$, a majority of rankers prefer $d_i$ to $d_j$, then the aggregate should prefer $d_i$ to $d_j$.}

We will focus on two classical rank aggregation techniques in this paper: Borda and Kemeny, describe below.
 
{\bf Borda Aggregation:}
In Borda aggregation~\cite{borda1781} each candidate is assigned a score by each ranker; where the score for a candidate is the number of candidates below him in each ranker's preferences. The Borda aggregation is the descending order arrangement of the average Borda score for each candidate averaged across all ranker preferences. Though Borda aggregation satisfies neutrality, monotonicity, and consistency, it does not satisfy the Condorcet Criterion~\cite{young78:siam} and ECC. In fact, it has been shown that no method that assigns weights to each position and then sorts the results by applying a function to the weights associated with each candidate satisfies the \ecc~\cite{dwork:www01}. This includes point-wise classifiers like \lr. This motivates us to consider order-based methods for rank aggregation that do satisfy ECC.

{\bf Kemeny Aggregation:}
A Kemeny optimal aggregation~\cite{kemeny59} is an aggregation that has the minimum number of pairwise disagreements with all rankers, i.e., a choice of $\tau$ that minimizes $K(\tau ,\tau _1 ,...,\tau _r ) = \frac{1}{r}\sum\limits_{i = 1}^r {k(} \tau ,\tau _i )$; where the function $k(\sigma,\tau)$ is the {\em Kendall tau} distance measured as $\left| {\{ (i,j)|i < j,\sigma (i) > \sigma (j),{\rm but }~\ \tau {\rm (i) < }\tau {\rm (j)}\} } \right|$, where $\sigma (i)$ 
is used to denote the position of $i$ in ranking $\sigma$. 

Kemeny aggregation satisfies neutrality, consistency, and the \ecc. Kemeny optimal aggregation also has a good maximum likelihood interpretation. Suppose there is an underlying ``correct'' ordering $\sigma$ of $S$, and each order $\tau_1,...,\tau_r$ is obtained from $\sigma$ by swapping pairs of elements with some probability less than $1/2$. That is, the $\tau$'s are ``noisy'' versions of $\sigma$. A Kemeny optimal aggregation of $\tau_1,...,\tau_r$ is one (not necessarily unique) that is maximally likely to have produced the $\tau$'s.

\section{Supervised Kemeny Ranking} 
\label{sec:suprankagg}

While Kemeny aggregation is optimal in the sense described above, it has two drawbacks when applied to our setting: (1) It is computationally very expensive, and (2) it does not distinguish between \textit{good} and \textit{bad} input rankings. Below we describe how we overcome these drawbacks. 

Kemeny (and Borda) aggregation, being motivated from Social Choice Theory, strive for \emph{fairness} and hence treat all rankers as equally important. However, fairness is not a desirable property in our setting, since we know that some individual rankers (measures) perform better than others in our target tasks. If we knew \textit{a priori} which rankers are better, we could leverage this information to produce a better aggregate ranking. In fact, given the ordering of a (small) set of candidates, we can estimate the performance of individual rankers and use this to produce a better ranking on a new set of candidates. We propose \skr (SKR), which is based on such an approach.

\drop{
There are two main challenges in implementing \skr: (1) Computation of optimal aggregation and (2) Supervision. In this section, we will first address the problem of computational hardness of optimal \skr, by proposing an \pk procedure. We then extend this procedure to \skr, by accomodating weigts to each ranker in the \pk method.
}

The problem of computing optimal Kemeny aggregation is NP-Hard for $r \ge 4$~\cite{dwork:www01}. However, there have been some attempts to approximately solve Kemeny optimal aggregation~\cite{frans09}. Ailon et al.~\cite{ailon08} presents a solution to the \emph{feedback arc set problem} on tournaments, which can be applied to rank aggregation for a 2-approximation of Kemeny optimal aggregation. We use this approach, which we refer to as \pk; and we show here that it satisfies a relaxation of Kemeny optimality and the \ecc. 

\pk can be described simply as a Quick Sort on elements based using the majority precedence relation $\succ$ as a comparator, where $d_i \succ d_j$ if the majority of input rankings has ranked $d_i$ before $d_j$. Note that, the relation $\succ$ is not transitive, and hence different comparison sort algorithms can produce different rankings. In \cite{dwork:www01} Dwork et al. propose the use of Bubble Sort, which also leads to an aggregation that satisfies ECC, but comes with no approximation guarantees. This approach, which they refer to as \lk, is one of the baselines in our experiments.  
\drop{For this reason, Local Kemnization does perform poorly in terms of both AUC and MAP compared to our Supervised Kemeny Aggregation algorithm.}

By extension from Quick Sort, it can be easily shown that \pk runs in $O(rm \log m)$. We show below that \pk also produces an aggregation that satisfies the following optimality criterion.

\begin{Definition}
\label{def:lk}
A permutation $\tau$ is {\em locally Kemeny optimal}~\cite{dwork:www01}, if there is no full list ${\tau}^+$ that can be obtained from $\tau$ by a single transposition of an adjacent pair of elements, such that, $K(\tau ^ +  ,\tau _1 ,...,\tau _r ) < K(\tau ,\tau _1 ,...,\tau _r )$.
\end{Definition}

\drop{In other words, local Kemeny optimality ensures that it is impossible to reduce the total Kendall tau distance between a permutation $\tau$ and all input rankers, by swapping an adjacent pair of elements in $\tau$.} 

\drop{
The quick sort algorithm proposed by Ailon et al. \cite{ailon08} is an expected 2-approximation algorithm. Using quick sort as the sorting mechanism in Local Kemnization takes us closer to Kemeny optimality. We call this as \pk algorithm as listed in Algo. \ref{algo:lk} except that the weights $w_i$ are always 1. The \pk procedure uses quick sorting mechanism, where given an initial ordering, elements $d_i$ is placed to left of $d_j$, if $d_i \succ d_j$ by the majority of rankers ($\tau_i$'s). We call our algorithm \pk, since the time taken to compute majority table $M$ is polynomial in $m$. However, using efficient data structures one can compute the $\tau$ in $O(rm \log m)$, where $m$ is the size of $V$. Using Lemma 1 and Theorem 1 we show that \pk procedure is local Kemeny optimal and satisfies ECC.
}

\begin{lemma}
\label{lemma:lk}
The final aggregation $\tau$ of the \pk procedure produces a locally optimal Kemeny order. 
\end{lemma}

\noindent\textsl{Proof}:
Every element in the final order is compared at least once with its neighboring elements in the quick sort procedure. As such, $d_i$ is placed immediately to the left of $d_j$ only if $d_i$ is preferred to $d_j$ by a majority of input rankings. So, swapping any such adjacent elements can only increase the number of input rankings that disagree with this ordering, thus increasing the total Kendall tau distance. Hence \pk is locally Kemeny optimal.
\hspace*{\fill}{$\blacksquare$}

\begin{theorem}
\label{theorem:ecc}
\drop{Let the final aggregation of \pk procedure be locally Kemeny optimal, then it satisfies ECC with respect to $\tau_1,\tau_2,...,\tau_r$.}
Let $\tau$ be the final aggregation of the \pk procedure. Then $\tau$ satisfies the \ecc with respect to the input rankings $\tau_1,\tau_2,...,\tau_r$. 
\end{theorem}

\noindent\textsl{Proof}:
The proof follows directly from Lemma 6 of \cite{dwork:www01}. 
If the claim is false then there exist rankers $\tau_1,\tau_2,...,\tau_r$, an \pk aggregation 
$\tau$, and a partition $(T,U)$ of the elements where for all $a \in T$ and $b \in U$ 
the majority among $\tau_1,\tau_2,...,\tau_r$ prefers $a$ over $b$, but there is a $c \in T$ and a $d \in U$ such that $d > c$ in $\tau$. 
Let $(d,c)$ be a closest such pair in $\tau$.
Consider the immediate successor of $d \in \tau$, and call it $e$. 
If $e=c$ then $c$ is adjacent to $d \in \tau$ and transposing this adjacent pair of elements produces a $\tau^+$ such that $K(\tau ^ +  ,\tau _1 ,...,\tau _r ) < K(\tau ,\tau _1 ,...,\tau _r )$, contradicting Lemma~\ref{lemma:lk} that $\tau$ is a locally Kemeny optimal aggregation of the $\tau_1,\tau_2,...,\tau_r$. 
If $e$ does not equal $c$, then either $e$ is in $T$, in which case the pair $(d,e)$ is a closer pair in $\tau$ than $(d,c)$ and also violates the \ecc, or $e$ is in $U$, in which case $(e,c)$ is a closer pair than $(d,c)$ that violates the \ecc. 
Both cases contradict the choice of $(d,c)$.
\hspace*{\fill}{$\blacksquare$}

The pseudo-code for \skr is presented in Algo.~\ref{algo:lk}. In order to accommodate supervision, we extend \ak aggregation to incorporate weights associated with each input ranking. The weights correspond to the relative utility of each ranker, which may depend on the task at hand. For the task of influence prediction in Twitter, we weigh each ranker based on its (normalized) AUC computed on a training set of candidates, for which we know the target variable i.e., the true retweet rates. When evaluating on \ap, we use weights based on \ap instead. For \skr we incorporate weights directly in sorting the elements through Quick Sort. Instead of comparing candidates based on the preference of the simple majority of individual rankers, we use a weighted majority. This can be achieved simply by using weighted votes during the creation of the majority table $M$ -- which represents the sum of weights of the rankers who prefer the row candidate to the column candidate for each pairwise comparison. 

Instead of using total orderings provided by each ranker, we can also use partial orderings (for a subset of candidates). Since identifying relevant candidates at the top of the list is usually more important, we use the partial orderings corresponding to the top $k$ candidates for each ranker. In our experiments, unless otherwise specified, we use the top-ranked $15\%$ of candidates for each ranker. 

\begin{algorithm}[tb]
\small
\caption{\skr (SKR)}
\label{algo:lk}
\begin{algorithmic}
\begin{STATE}
{\bf Input:}
$\tau_i=[\tau_{i1},...,\tau_{im}],\forall i=1,...,r$, ordered arrangement of m candidates for r rankers.\\
$w=[w_1,...,w_r]$ -- where $w_i$ is the weight of ranker $i$\\
$\mu=[\mu_1,...,\mu_m]$ -- initial ordered arrangement of m candidates\\
$k$ -- the number candidates to consider in each ranker's preference list ($k \le m$)\\
{\bf Output:} 
$\tau$ -- rank aggregated arrangement of candidates in decreasing order of importance\\
\vspace{.03in}
\begin{enumerate}
\item{Initialize majority table $M_{i,j} \Leftarrow 0,\forall i,j=1,...,m$}
\vspace{-0.01in}
\item{For each ranker $p = 1$ to r}\\
\vspace{-0.01in}
\item{\hspace{0.2in}For each candidate $i = 1$ to k-1}\\
\vspace{-0.01in}
\item{\hspace{0.4in}For each candidate $j = i+1$ to k}\\
\vspace{-0.01in}
\item{\hspace{0.6in}$M_{\tau_{pi},\tau_{pj}} \Leftarrow M_{\tau_{pi},\tau_{pj}}+w_p$}\\
\vspace{-0.01in}
\item{\label{step:sort} Quick sort $\mu$, using $M_{\mu_i,\mu_j}$. If $M_{\mu_i,\mu_j}-M_{\mu_j,\mu_i} > 0$ then $\mu_i$ is greater than $\mu_j$. If $M_{\mu_i,\mu_j}-M_{\mu_j,\mu_i} = 0$ then $\mu_i$ is equal to $\mu_j$. If $M_{\mu_i,\mu_j}-M_{\mu_j,\mu_i} < 0$ then $\mu_i$ is less than $\mu_j$.}\\
\vspace{-0.01in}
\item{Return $\tau$}
\vspace{-0.01in}
\end{enumerate}
\end{STATE}
\end{algorithmic}
\end{algorithm}

\section{Empirical Evaluation}

We compared \skr to using individual rankings, \lr using all input rank scores as features, 
\lk~\cite{dwork:www01}, Borda aggregation, and a supervised version of Borda aggregation. We also  compared to SVMRank~\cite{joachims:06}, which is a supervised approach that tries to optimize performance on AUC. 

For \supborda, we incorporate performance-based (AUC/AP) weights in Borda aggregation\drop{, as presented in Algo.~\ref{algo:borda}}. This is relatively straightforward, where instead of simple averages, we take weighted averages of Borda scores. A similar approach to supervised Borda was used in \cite{aslam01:sigir}, where weights were based on average precision of each ranker for a meta-search task. While, supervised versions of Borda appear in prior work, to our knowledge, we present the first supervised version of Kemeny aggregation.\footnote{A very preliminary version of our work appears in \cite{melville:win10}}

\drop{
\begin{algorithm}[tb]
\small
\caption{Supervised Borda}
\label{algo:borda}
\begin{algorithmic}
\begin{STATE}

{\bf Input:} 
$\tau_i=[\tau_{i1},...,\tau_{im}],\forall i=1,...,r$, ordered arrangement of $m$ candidates for $r$ rankers.\\
$w=[w_1,...,w_r]$ -- where $w_i$ is the weight for ranker $i$\\
{\bf Output:} 
$\tau$ -- rank aggregated arrangement of candidates in decreasing order of importance\\
\vspace{.08in}
\begin{enumerate}
\item{Initialize $\beta_i \Leftarrow 0,\forall i=1,...,m$, $\beta_i$ is the borda score of candidate $i$}
\vspace{-0.01in}
\item{For each ranker $p = 1$ to r}\\
\vspace{-0.01in}
\item{\hspace{0.2in}For each candidate $i = 1$ to m}\\
\vspace{-0.01in}
\item{\hspace{0.4in}$\beta_{i} \Leftarrow \beta_i + ((m-i+1)w_p)$}\\
\vspace{-0.01in}
\item{Sort $\beta_i$ in descending order, such that $\tau=[d_1,...,d_m],\beta_{d_i} \geq \beta_{d_j},\forall i,j=1,...,m, i \neq j$}
\vspace{-0.01in}
\item{Return $\tau$}
\end{enumerate}
\end{STATE}
\end{algorithmic}
\end{algorithm}
}

In order to verify the effectiveness of each component of \skr, we performed several ablation studies. In particular, we compared \skr to the following variations of Algo.~\ref{algo:lk}:
\begin{itemize}
\item \textit{Unsupervised, Total Orderings}: Using uniform weights ($w_i = 1, \forall i)$, and $k = |S|$, which reduces to the unsupervised approximation to Kemeny aggregation on total orderings.
\item \textit{Supervised, Total Orderings}: $k = |S|$, i.e., \skr on total orderings.
\item \textit{Unsupervised, Partial Orderings}: Using uniform weights ($w_i = 1, \forall i)$.
\item \textit{Supervised, Bubble Sort}: Using Bubble Sort instead of Quick Sort in Step~\ref{step:sort}. This can be viewed as a supervised version of \lk~\cite{dwork:www01}.
\end{itemize}

\subsection{Twitter Network Study}

We compared our approach, \skr, to the different supervised and unsupervised techniques described above on the task of predicting viral potential, as in Sec.~\ref{sec:data}. As inputs to each aggregation method we use the 13 different measures listed in Table~\ref{table:measures}. Each measure is used to produce a total ordering of preferences over the 9,625 candidates (twitter users), where ties are broken randomly. We compared the 10  aggregation methods (see Table~\ref{table:rankAgg}) to individual rankers, but in the interest of space we only list the best individual measure (\rtgwtdindegree) in the table. We averaged performance, measured by AUC and Average Precision@100, over 10 runs of random stratified train-test splits for different amounts of data used for training. These results are summarized in Tables~\ref{table:rankAgg} and~\ref{table:rankAggMAP}.

We note that, in terms of AUC, in general, aggregation techniques perform better than using \rtgwtdindegree, which is the best individual ranker. However, apart from \skr, this is not always the case for Average Precision. So one must use rank aggregation with caution, depending on the desired performance metric. The results also show that our version of \supborda performs better than traditional Borda aggregation. However, \lk, outperforms \supborda, showing the benefit of Kemeny-based aggregation versus Borda's score-based aggregation. Our approach, of \skr, further improves on this result, with the best performance at all points in terms of Average Precision, and 3 of 4 points in terms of AUC. Logistic Regression is a little better than \skr at one point in terms of AUC. However, overall \lr is less effective than the other aggregation methods, occasionally performing worse than the best individual ranker.  
\skr, also outperforms SVMRank, consistently on all training sample sizes, in both AUC and AP.\footnote{Note that, while some absolute differences may appear small, a relative improvement of 1\% is considered to be substantial in ranking domains such as web search (see Fig. 1 of \cite{zheng:nips07}).}

\drop{\footnote{Improvements reported in web search ranking papers are typically of the order of 1\%; see \drop{for instance} Fig. 1 of \cite{zheng:nips07}}.} 

\drop{
Our proposed method, \skr, outperforms learning to rank methods like SVM Rank, consistently on all training sample sizes in both AUC and AP performance measures. Note that a relative improvement of 1\% is considered to be substantial in the ranking domain. \footnote{What we mean here is that improvements reported in the ranking papers are typically of the order of 1\%; see for instance Fig. 1 of \cite{zheng:nips07}}. In addition, \skr consistently outperforms existing state-of-the-art influence measures used in popular websites like Twitterholic.com (Followers) and Tunrank.com (follower pagerank), as shown in Fig. \ref{fig:viral-roc}.
}

Our ablations studies show that every component of \skr does contribute to its superior performance. In particular, we see that supervised variants of Algo.~\ref{algo:lk} perform better than unsupervised variants. Also, focusing on the top $k$ elements from each individual ranker (\textit{partial orderings}) is more effective that using total orderings. Finally, using the Quick Sort approximation to Kemeny aggregation makes a notable difference over using Bubble Sort. As mentioned earlier, the Bubble Sort variation, as used by Dwork et al.~\cite{dwork:www01} comes with no approximation guarantees, which makes a perceptible difference in practice. In addition to using AUC-based weights for \skr, we also experimented with alternative weighting schemes in Algo.~\ref{algo:lk}, such as, $(AUC-0.5)$ and ($log(AUC/(1-AUC))$). However, in experiments (not presented) the simple AUC based weights outperformed other weighting schemes by a margin of $2-5\%$. \drop{Selecting the optimal weights requires solving a computational infeasible quadratic program.}

\drop{
In addition, we also tested various other weighting schemes, like, AUC offset by 0.5 $(AUC-0.5)$ and sigmoid weighting ($log(AUC/1-AUC)$) and Shapley value approximation \cite{narayanam08}. The Shapley value approach is a well-known mechanism in cooperative game theory, for the division of surplus amongst a coalition of users. In our weighting scheme, Shapley value is used to determine marginal contribution of individual rankers when they are combined. Out of these different schemes, we found that, the results presented in this paper using the simple AUC based weights outperformed all other weighting schemes by a margin of 2-7\%. Also, our proposed method can take any other influence models \cite{ghosh:snakdd10}, and learning to rank methods (like SVMRank, RankBoost, etc), in addition to what we have already described as input to the rank aggregation method, to further improve the performance in terms of AUC and AP.
}

\begin{figure}[tbh]
\centering
\includegraphics[scale=0.5]{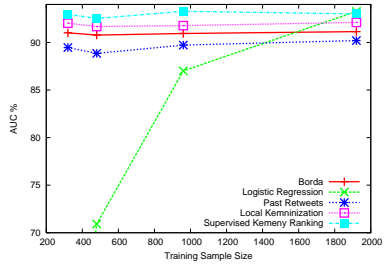}
\caption{AUC performance of rank aggregation techniques with increasing training data.}
\label{fig:rankAgg}
\end{figure}

\begin{figure}[tbh]
\centering
\includegraphics[scale=0.38]{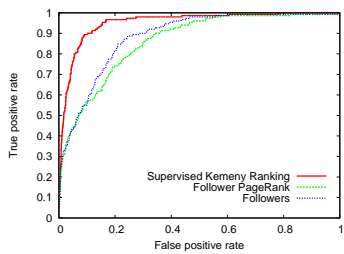}
\caption{ROC curves comparing \skr to popular measures in practice.}
\label{fig:viral-roc}
\end{figure}

Learning curves comparing our approach to existing baselines are presented in Fig.~\ref{fig:rankAgg}. We observe that, while \lr performs well with ground truth on a large number of candidates, its performance drops significantly with lower levels of supervision. In contrast, the rank aggregation methods are fairly stable, consistently beating the best individual ranking and performing better than \lr in the more realistic setting of moderately-sized training sets. The consistently good performance of \skr confirms the advantages of supervised locally optimal order-based ranking compared to score-based aggregation, such as \borda, and unsupervised methods. 

While Fig.~\ref{fig:rankAgg} shows the performance in terms of area under the ROC curve for different sample sizes, in Fig.~\ref{fig:viral-roc} we present the ROC curves for a single point (1,920 training samples). We contrast \skr, with the methods most commonly used in practice, namely, number of followers and follower PageRank (e.g., as done by Twitaholic.com and Tunrank.com). Note that, all other baselines in this paper are devised by us, and are much better than these approaches. We observe that \skr performs 5 to 8\% better in terms of AUC and 54 to 55\% better in terms of AP compared to current practice.

\begin{table}[tb]
\tiny
\begin{center}
\begin{tabular}{|p{2.9cm}|p{0.75cm}|p{0.75cm}|p{0.75cm}|p{0.75cm}|}
\hline
\textbf{}&\multicolumn{4}{|c|}{\textbf{Training Samples}}\\
\hline
\textbf{Ranking Method}&\textbf{320}&\textbf{480}&\textbf{960}&\textbf{1920}\\
\hline
\hline

\skr & \textbf{92.97}	& \textbf{92.52}	& \textbf{93.28}	& 93.00\\
\rtgwtdindegree &89.47	& 88.86 & 89.73	& 90.20\\
\lr & 46.87	& 70.92 & 87.02 & \textbf{93.26}\\
\bordaunsup & 91.02 & 90.78 & 90.95 & 91.14\\
\bordasup & 91.50 & 91.09	& 91.22 & 91.62\\
Local Keminization & 92.03 & 91.68 & 91.78 & 92.11\\
SVM Rank & 87.98 & 89.33 & 92.15 & 92.79 \\
\hline
\hline

\lkunsupborda & 88.49	& 88.29	& 89.91	& 89.35 \\
\lksupborda & 88.89 &	88.36	& 89.92	& 89.51\\
\lktkunsupbordasup & 92.73	& 92.42	& 92.72	& 92.58\\
Supervised, Bubble Sort & 92.23 & 91.88 & 92.03 & 92.27 \\

\hline
\end{tabular}
\end{center}
\caption{\label{table:rankAgg}Rank aggregation performance measured in AUC(\%) for various training set sizes.}
\end{table}

\begin{table}[tb]
\tiny
\begin{center}
\begin{tabular}{|p{2.9cm}|p{0.75cm}|p{0.75cm}|p{0.75cm}|p{0.75cm}|}
\hline
\textbf{}&\multicolumn{4}{|c|}{\textbf{Training Samples}}\\
\hline
\textbf{Ranking Method}&\textbf{320}&\textbf{480}&\textbf{960}&\textbf{1920}\\
\hline
\hline

\skr & \textbf{0.7242} & \textbf{0.6837} & \textbf{0.6991} & \textbf{0.6783}\\
\rtgwtdindegree & 0.7210 & 0.6610 & 0.6766 & 0.6668\\
\lr & 0.3255 & 0.4862 & 0.6662 & 0.6219\\
\bordaunsup & 0.2600 & 0.2600 & 0.2333 & 0.2133\\
\bordasup & 0.3000 & 0.2733 & 0.2366 & 0.2334\\
Local Keminization & 0.5240 & 0.4938 & 0.4768 & 0.4891\\
SVM Rank & 0.1732 & 0.3180 & 0.3990 & 0.3996 \\
\hline
\hline

\lkunsupborda & 0.6982 & 0.5998 & 0.6706 & 0.6357\\
\lksupborda & 0.6994 & 0.6024 & 0.6826 & 0.6521\\
\lktkunsupbordasup & 0.7018 & 0.6622 & 0.6745 & 0.6619\\
Supervised, Bubble Sort & 0.5273 & 0.4963 & 0.4772 & 0.4930\\
\hline
\end{tabular}
\end{center}
\caption{\label{table:rankAggMAP}Rank aggregation performance measured in Average Precision@100 for various training set sizes.}
\end{table}

\subsection{Citation Network Study}
\label{sec:citation}

In addition to Twitter data, we also performed a case study on publication citation networks. For this we used a collection of papers with their citations that was used in the KDD Cup contest held in 2003.\footnote{http://www.cs.cornell.edu/projects/kddcup/} This data  consists of 1,716 papers in the field of High Energy Physics Theory (hep-th), published on arXiv.org during a 6 month period. The data set also contains the number of times each paper was downloaded during the 60 day period after it was published on arXiv.org. This download information gives us an extrinsic proxy for the influence of a paper. As such, we define the task of predicting highly influential papers, as measured by downloads, based on the citation data of the papers. If a paper received 600 or more downloads, we consider it as a high-influence paper (77 papers); else we consider it to have little or no influence.

First, we constructed a citation graph based on all publications in hep-th, which was also provided as part of KDD Cup 2003. In this citation graph, each node represents a paper and each edge represents a citation. As of May 1, 2003, there were 29,014 papers and 342,427 citations in total in the hep-th data. Next, for each of the 1,716 papers with download information, we used this citation graph to compute 5 influence measures - Indegree, Outdegree, Pagerank, Hub and Authority score \cite{kleinberg:acm99}.
 
We ran experiments as before, using 20\% of the data (343 papers) for training the supervised methods, and setting $k$ to 1,200 in Algo.~\ref{algo:lk}. The results in terms of AUC and Average Precision for each method are presented in Table~\ref{table:citmeasures}. As expected, the number of papers citing a given paper (in-degree) is a good indicator of how often the paper will be downloaded. Furthermore, having more citations from highly-cited papers, as captured by PageRank is a better indicator of influence in this data. Note that, this was not the case in predicting viral potential in Twitter. The number of papers a paper is citing (out-degree) and Hub-score have some, though weaker, ability to predict influence. This is probably because some survey papers do become influential if they refer to many good papers in that area. 

\begin{table}[tbh]
\tiny
\begin{center}
\begin{tabular}{|l|l|l|}
\hline
\textbf{Measure}&\textbf{AUC \%}&\textbf{AP}\\
\hline
\hline
\pr	&81.09&0.4470\\
Indegree	&80.42&\textbf{0.5376}\\
Authority &80.39&0.5324\\
Outdegree &64.33&0.2820\\
Hub 	&61.07&0.2867\\
\hline
\hline
\skr &\textbf{81.70}& 0.4950\\
\lr	&76.02&0.5330\\
\borda &77.47&0.2363\\
\supborda &78.27&0.2787\\
Local Keminization &76.62&0.3668\\
SVMRank &77.59&0.4625\\
\hline
\hline
\lkunsupborda &80.12&0.3518\\
\lksupborda &80.30&0.4902\\
\lktkunsupbordasup & 80.23 & 0.4928 \\
Supervised, Bubble Sort & 79.17 & 0.4798\\
\hline
\end{tabular}
\end{center}
\caption{\label{table:citmeasures} Comparing ranking methods for identifying influential papers, based on AUC and Average Precision@60.}
\end{table}

In this study we find that not all aggregation techniques are better than using individual rankers. In particular, high in-degree is very correlated with high download rates, as reflected by Average Precision. So depending on the data and the evaluation metrics, one should always consider using the best individual ranker along with alternative aggregation methods. Nevertheless, in terms of AUC, \skr still produces the best ranking, outperforming individual rankers and other aggregation techniques. \drop{In addition to the baselines we used for the Twitter data, here, the dataset was small enough to run SVMRank. While it performed comparably to other baselines, SVMRank is still outperformed by \skr.} The results on the ablation studies are similar to before, further corroborating the contribution of each component of the \skr algorithm.   
 
\drop{
These results also substantiate the fact that supervised rank aggregation algorithms perform better than their unsupervised counterparts. Notably, \skr outperforms all individual measures, rank aggregation techniques and prior art like SVMRank and \lk.
\prem{Update based on AP results.}
}

\section{Related work}
\label{sec:related}

\drop{There has been a lot of recent work studying influence and the diffusion of information in social networks~\cite{bakshy:ec09,goetz:icwsm09,kossinets:kdd08}. \drop{These works are very helpful to understand the dynamics of communication in networks, and allows us ways of comparing different kinds of communication within different network.} However, they do not directly give us measures of influence and authority in graphs.} \drop{It would be useful to follow the analyses done in these papers to derive something actionable, such as a better measure of influence in social media, or insight into how to increase reach in such domains.} An associated growing area of research attempts to explain content and link structures in social media, together with their temporal evolution, based on tensor factorizations and higher order extensions of techniques such as Singular Value Decomposition (SVD)~\cite{kolda:sdm-wkshp06,chi:kdd07}. Recently, Weng et al.~\cite{weng:wsdm10} propose TwitterRank, a variant of \pr that also takes topical similarity between users into account.

Another interesting approach to quantitatively evaluating the ranking of blogs is through the task of cascade detection –- selecting a set of blogs to read which link to most of the stories that propagate over the blogosphere. Current solutions~\cite{leskovec:kdd07,kempe:kdd03} to this task do not attempt to address the task of assigning an influence score to individual bloggers, since they are focused on optimal set selection. However, there is a lot of potential for using such approaches to identify influencers.

In related work on rank aggregation, Liu et al.~\cite{liu:www07} present an alternative supervised approach for the task of web-search -- where they build on a Markov Chain (MC) based approach to rank aggregation. However, it has been shown that \lk improves on MC-based approaches~\cite{dwork:www01}, which in turn, we show is outperformed by \skr.
 
In concurrent work on the analysis of Twitter, Cha et al.~\cite{cha:icwsm10} also conclude that number of followers alone reveals little about a user's influence. We go further in our work, by comparing many more socio-metrics on different tasks, and providing approaches to improve influence prediction through rank aggregation. In recent work, Suh et al.~\cite{suh:sc2010} analyze factors that correlate with retweeting. While they consider in- and out-degrees of the follower graph, they do not look at other graphs, such as the retweet graph, or other socio-metrics, such as PageRank. Furthermore, since their study only uses randomly sampled tweets, they are limited to a very small subset of retweets. In contrast, we collect all retweets for all users in our study.

In addition to SVMRank, there have been several recent advances in learning to rank~\cite{freund:jmlr03,burges:jmlr11}, driven largely by the application to web search. All of these approaches produce a ranked list as an output. In their seminal work, Dwork et al.~\cite{dwork:www01}, showed how rank aggregation can be used to improve on meta-search, by combining individual search rankings. Since, we demonstrate that \skr performs better than their \lk approach, we are hopeful that it can be used to aggregate the rankings from different learning to rank methods, to improve results on web search and other applications.

In recent work, Ghosh and Lerman~\cite{ghosh:snakdd10} evaluate various influence models based on geodesic-path based distance measures and topological ranking measures. They propose a Normalized $\alpha$-centrality algorithm and evaluate its effectiveness on measuring influential users in Digg.com. Their work aims to find the best individual socio-metric and does not intend to improve the predictive accuracy by combining  various influence models. However, as we have shown in this paper, often individual socio-metrics fail to capture all critical factors that are relevant for predicting influence in networks. Presumably, one could use the Normalized $\alpha$-centrality algorithm as another input ranker to \skr, to further improve predictive performance.

The work by Agarwal et al.~\cite{agarwal:wsdm08} does a \drop{thorough} empirical study 
on identifying influential people in blog networks. They propose 4 main features that 
produce influence in the bloggers network, based on recognition, activity, 
novelty, and eloquence. They weigh these four features to produce a combined 
score for each blogger. 
In \cite{sayyadi:sdm09}, Sayyadi and Getoor predict the popularity of a paper 
using its expected future citations. They propose $FutureRank$, which combines the PageRank score of 
a paper in the citation network, the authority score in the authorship network, and 
the recency of the publication. 
Both \cite{agarwal:wsdm08} and \cite{sayyadi:sdm09} propose a score-based model, 
where they combine the scores from a set of features defined on the underlying 
network data. Note that, neither of the methods are supervised and they require 
further enhancements to accommodate such supervision. In addition, their methods 
are score-based aggregations, and not order-based. Both Dwork et al.~\cite
{dwork:www01} and this paper shows clearly the inefficiency of weighted 
combination of score-based algorithms compared to order-based. 

\section{Conclusion and Future Work}
\label{sec:conclusions}

\drop{The analysis of social media is revolutionizing the way we think about actor importance and centralities in social networks.} 

Understanding influence within blog and micro-blog networks has become a crucial technical problem with increasing relevance to marketing and information retrieval. We address the problem of assessing influence by casting it in the form of a predictive task; which allows us to objectively compare different measures of influence in light of standard classification and ranking metrics. Furthermore, we propose a novel supervised rank aggregation method, which combines aspects of different influence measures to produce a composite ranking mechanism that is most effective for the desired task. We have applied this approach to a case study involving 40 million twitter accounts, and have examined the task of predicting the potential for viral out-breaks. We further corroborated these results on the task of identifying influential papers based on citation networks. Empirical results show that our proposed approach, \skr, performs better than several existing rank aggregation techniques, as well as other supervised learning benchmarks. 

The problem of choosing the optimal Kemeny order can be formulated as a mixed-integer programming problem as discussed in~\cite{conitzer:aaai06}. However, the problem of finding the optimal weights for \skr is much more difficult, as it involves a quadratic objective function, with two sets of variables; one for selecting the optimal weights and one for the optimal order. An efficient algorithm to solve this optimization could significantly improve results, and is a promising direction for future work.


\drop{
This paper makes two main contributions in that direction.
We also demonstrated the merits of supervised locally Kemeny optimal rank aggregation. 
\drop{Hopefully, this study will motivate the quantitative comparison and creation of more task-driven socio-metrics.}
}

\section*{Acknowledgements}
We would like to thank Estepan Meliksetian for the help in gathering the Twitter data set. We are also grateful to Claudia Perlich, Richard Lawrence and Andrew Davenport for their suggestions and comments on this work.

\bibliographystyle{IEEEtran}
\bibliography{influence,banter} 

\begin{thebibliography}{10}
\providecommand{\url}[1]{#1}
\csname url@samestyle\endcsname
\providecommand{\newblock}{\relax}
\providecommand{\bibinfo}[2]{#2}
\providecommand{\BIBentrySTDinterwordspacing}{\spaceskip=0pt\relax}
\providecommand{\BIBentryALTinterwordstretchfactor}{4}
\providecommand{\BIBentryALTinterwordspacing}{\spaceskip=\fontdimen2\font plus
\BIBentryALTinterwordstretchfactor\fontdimen3\font minus
  \fontdimen4\font\relax}
\providecommand{\BIBforeignlanguage}[2]{{%
\expandafter\ifx\csname l@#1\endcsname\relax
\typeout{** WARNING: IEEEtran.bst: No hyphenation pattern has been}%
\typeout{** loaded for the language `#1'. Using the pattern for}%
\typeout{** the default language instead.}%
\else
\language=\csname l@#1\endcsname
\fi
#2}}
\providecommand{\BIBdecl}{\relax}
\BIBdecl

\bibitem{bakshy:ec09}
E.~Bakshy, B.~Karrer, and L.~Adamic, ``{Social influence and the diffusion of
  user-created content},'' in \emph{ACM EC}, 2009.

\bibitem{goetz:icwsm09}
M.~Goetz, J.~Leskovec, M.~Mcglohon, and C.~Faloutsos, ``{Modeling Blog
  Dynamics},'' in \emph{ICWSM}, 2009.

\bibitem{kossinets:kdd08}
G.~Kossinets, J.~Kleinberg, and D.~Watts, ``{The structure of information
  pathways in a social communication network},'' in \emph{KDD}, 2008.

\bibitem{moreno:book34}
J.~Moreno, \emph{{Who Shall Survive? Foundations of Sociometry, Group
  Psychotherapy and Sociodrama}}.\hskip 1em plus 0.5em minus 0.4em\relax
  Nervous and Mental Disease Publishing Co., 1934.

\bibitem{knoke:ana83}
D.~Knoke and R.~Burt, \emph{{Applied Network Analysis}}.\hskip 1em plus 0.5em
  minus 0.4em\relax Newbury Park, CA: Sage, 1983, ch. Prominence.

\bibitem{wasserman:book94}
S.~Wasserman and K.~Faust, \emph{{Social Network Analysis: Methods \&
  Applications}}.\hskip 1em plus 0.5em minus 0.4em\relax Cambridge, UK:
  Cambridge University Press, 1994.

\bibitem{borda1781}
J.~Borda, ``Memoire sur les elections au scrutin,'' in \emph{Histoire de
  l'Academie Royale des Sciences}, 1781.

\bibitem{kemeny59}
J.~Kemeny, ``Mathematics without numbers,'' in \emph{Daedalus}, vol.~88, 1959,
  pp. 571--–591.

\bibitem{faw:prl06}
T.~Fawcett, ``An introduction to roc analysis,'' in \emph{Pattern Recognition
  Letters}, vol.~27, 2006, pp. 861--874.

\bibitem{yates:99}
R.~Baeza-Yates and B.~Ribeiro-Neto, ``Modern information retrieval.''\hskip 1em
  plus 0.5em minus 0.4em\relax Addison Wesley Co, 1999.

\bibitem{arrow63}
K.~Arrow, ``Social choice and individual values.''\hskip 1em plus 0.5em minus
  0.4em\relax New Haven: Cowles Foundation, 2nd Edition 1963.

\bibitem{truchon98}
M.~Truchon, ``An extension of the condorcet criterion and kemeny orders,'' in
  \emph{J. Eco. Lit.}, 1998.

\bibitem{young78:siam}
H.~Young and A.~Levenglick, ``A consistent extension of condorcet's election
  principle,'' in \emph{SIAM J. on App. Math}, vol. 35(2), 1978.

\bibitem{dwork:www01}
C.~Dwork, R.~Kumar, R.~Naor, and D.~Sivakumar, ``Rank aggregation methods for
  the web,'' in \emph{WWW}, 2001.

\bibitem{frans09}
F.~Schalekamp and A.~van Zuylen, ``Rank aggregation: Together we're strong,''
  in \emph{ALENEX}, 2009, pp. 38--51.

\bibitem{ailon08}
N.~Ailon, M.~Charikar, and A.~Newman, ``Aggregating inconsistent information:
  Ranking and clustering,'' \emph{J. ACM}, vol.~55, no.~5, 2008.

\bibitem{joachims:06}
T.~Joachims, ``Training linear svms in linear time,'' in \emph{KDD}, 2006.

\bibitem{aslam01:sigir}
J.~A. Aslam and M.~Montague, ``Models for metasearch,'' in \emph{SIGIR}, 2001.

\bibitem{melville:win10}
P.~Melville, K.~Subbian, C.~Perlich, R.~Lawrence, and E.~Meliksetian, ``A
  predictive perspective on measures of influence in networks,'' in
  \emph{Proceedings of the Workshop on Information in Networks}, 2010.

\bibitem{zheng:nips07}
Z.~Zheng, H.~Zha, T.~Zhang, O.~Chapelle, K.~Chen, and G.~Sun, ``A general
  boosting method and its application to learning ranking functions for web
  search,'' in \emph{NIPS}, 2007.

\bibitem{kleinberg:acm99}
J.~Kleinberg, ``Authoritative sources in a hyperlinked environment,'' in
  \emph{J. ACM}, 1999.

\bibitem{kolda:sdm-wkshp06}
T.~Kolda and B.~Bader, ``{The TOPHITS model for higher-order web link
  analysis},'' in \emph{SDM Workshop on Link Analysis, Counterterrorism and
  Security}, 2006.

\bibitem{chi:kdd07}
Y.~Chi, S.~Zhu, X.~Song, J.~Tatemura, and B.~Tseng, ``Structural and temporal
  analysis of the blogosphere through community factorization,'' in \emph{KDD},
  2007.

\bibitem{weng:wsdm10}
J.~Weng, E.-P. Lim, J.~Jiang, and Q.~He, ``{T}witter{R}ank: Finding
  topic-sensitive influential {T}witterers,'' in \emph{WSDM}, 2010.

\bibitem{leskovec:kdd07}
J.~Leskovec, A.~Krause, C.~Guestrin, C.~Faloutsos, J.~Vanbriesen, and
  N.~Glance, ``{A Cost-effective outbreak detection in networks},'' \emph{KDD},
  2007.

\bibitem{kempe:kdd03}
D.~Kempe, J.~Kleinberg, and E.~Tardos, ``{Maximizing the spread of influence
  through a social network},'' in \emph{KDD}, 2003.

\bibitem{liu:www07}
Y.-T. Liu, T.-Y. Liu, T.~Qin, and H.~Li, ``Supervised rank aggregation,'' in
  \emph{WWW}, 2007.

\bibitem{cha:icwsm10}
M.~Cha, H.~Haddadi, F.~Benevenuto, and K.~Gummadi, ``Measuring user influence
  in twitter: The million follower fallacy,'' in \emph{ICWSM}, 2010.

\bibitem{suh:sc2010}
B.~Suh, L.~Hong, P.~Pirolli, and E.~H. Chi, ``Want to be retweeted? large scale
  analytics on factors impacting retweet in twitter network,'' in
  \emph{SocialCom}, 2010.

\bibitem{freund:jmlr03}
Y.~Freund, R.~Iyer, R.~Schapire, and Y.~Singer, ``An efficient boosting
  algorithm for combining preferences,'' in \emph{JMLR}, 2003.

\bibitem{burges:jmlr11}
C.~J.~C. Burges, K.~M. Svore, P.~N. Bennett, A.~Pastusiak, and Q.~Wu,
  ``Learning to rank using an ensemble of lambda-gradient models,'' in
  \emph{JMLR}, 2011, pp. 253--35.

\bibitem{ghosh:snakdd10}
R.~Ghosh and K.~Lerman, ``Predicting influential users in online social
  networks,'' in \emph{SNA-KDD}, 2010.

\bibitem{agarwal:wsdm08}
N.~Agarwal, H.~Liu, L.~Tang, and P.~S. Yu, ``Identifying the influential
  bloggers in a community,'' in \emph{WSDM}, 2008.

\bibitem{sayyadi:sdm09}
H.~Sayyadi and L.~Getoor, ``Futurerank: Ranking scientific articles by
  predicting their future pagerank,'' in \emph{SDM}, 2009.

\bibitem{conitzer:aaai06}
V.~Conitzer, A.~J. Davenport, and J.~Kalagnanam, ``Improved bounds for
  computing kemeny rankings,'' in \emph{AAAI}, 2006.

\end{thebibliography}

\end{document}